\documentclass[12pt,a4paper]{article}
\usepackage{a4wide}
\usepackage{slashed}
\usepackage{latexsym}
\usepackage{epsf}
\usepackage{amssymb}
\makeatletter
\@addtoreset{equation}{section}
\makeatother

\begin{document}
\begin{flushright}
\small
IFT-UAM/CSIC-02-17\\
SISSA 49/02/EP\\
{\bf hep-th/0206200}\\
June $21$st, $2002$
\normalsize
\end{flushright}

\begin{center}


\vspace{.7cm}

{\Large {\bf On $d=4,5,6$ Vacua with 8 Supercharges}}

\vspace{.7cm}


{\bf\large Ernesto Lozano-Tellechea}${}^{\spadesuit\heartsuit}$
\footnote{E-mail: {\tt Ernesto.Lozano@uam.es}},
{\bf\large Patrick Meessen}${}^{\aleph}$
\footnote{E-mail: {\tt meessen@sissa.it}}\\~ \\
{\bf\large and Tom{\'a}s Ort\'{\i}n}${}^{\spadesuit\clubsuit}$
\footnote{E-mail: {\tt tomas@leonidas.imaff.csic.es}}
\vskip 1truecm

${}^{\spadesuit}$\ {\it Instituto de F\'{\i}sica Te{\'o}rica, C-XVI,
Universidad Aut{\'o}noma de Madrid\\
Cantoblanco, E-28049 Madrid, Spain}

\vskip 0.2cm
${}^{\heartsuit}$\ {\it Departamento de F\'{\i}sica Te{\'o}rica, C-XI,
Universidad Aut{\'o}noma de Madrid\\
Cantoblanco, E-28049 Madrid, Spain}

\vskip 0.2cm
${}^{\aleph}$\ {\it International School for 
Advanced Studies (SISSA/ISAS)\\
Via Beirut 2-4, I-34014 Trieste, Italy}


\vskip 0.2cm
${}^{\clubsuit}$\ {\it I.M.A.F.F., C.S.I.C., 
Calle de Serrano 113 bis\\ 
E-28006 Madrid, Spain}
\vspace{.7cm}

{\bf Abstract}
\end{center}
\begin{quotation}
\small
We show how all known $N=2, d=4,5,6$ maximally supersymmetric vacua
(H$pp$-waves and $aDS\times S$ solutions) are related through
dimensional reduction/oxidation preserving all the unbroken
supersymmetries.  In particular we show how the $N=2, d=5$ family of
vacua (which are the near-horizon geometry of supersymmetric rotating
black holes) interpolates between $aDS_{2}\times S^{3}$ and
$aDS_{3}\times S^{2}$ in parameter space and how it can be
dimensionally reduced to an $N=2, d=4$ dyonic Robinson-Bertotti
solution with geometry $aDS_{2}\times S^{2}$ and oxidized to an $N=2,
d=6$ solution with $aDS_{3} \times S^{3}$ geometry (which is the
near-horizon limit of the self-dual string).
\end{quotation}
\newpage
\pagestyle{plain}
\section*{Introduction}
There is currently a renewed interest on maximally supersymmetric
vacua stemming from the discovery, and re-discovery of previously
overlooked, maximally supersymmetric H$pp$-wave solutions
\cite{Kowalski-Glikman:wv,Figueroa-O'Farrill:2001nz,Blau:2001ne}.
These solutions have very interesting properties: they are not only
supergravity solutions (i.e.~solutions of the lowest-order superstring
effective action) but it can be argued that they are exact solutions
of superstring theory to all orders and therefore good vacua on which
superstrings can be quantized \cite{Gueven:ad,art:RR_quant} and the
D-branes can be discussed \cite{art:D_branes,Skenderis:2002vf}. 
Further, these solutions can be obtained by a limiting procedure that
preserves (or increases) the number of unbroken (super)symmetries
\cite{kn:Pen6,Gueven:2000ru,Blau:2002dy} (for a review see,
e.g.~Ref.~\cite{Blau:2002rg}), a feature which has given rise to the 
Hpp/CFT correspondence (See {\em e.g.} \cite{art:hpp_cft}).
\par 
It is the standard lore that maximally supersymmetric vacua (other
than products of Minkowski spacetime by circles) of higher-dimensional
supergravity theories cannot be dimensionally reduced preserving all
their unbroken supersymmetries (See e.g. \cite{Duff:1998us,Michelson:2002wa}
and references therein): in general, the Killing spinors of
these vacua depend on all coordinates.  This dependence complicates
its compactification and dimensional reduction. First, only for
certain radii of the compact direction the Killing spinors will have
the right periodicity and, thus, only for those radii the compactified
solutions preserve the same amount of supersymmetry as the
uncompactified one. Second, unless the Killing spinors are independent
of the compact coordinates (or have a very special dependence on them,
as in some generalized dimensional reductions
\cite{Gheerardyn:2001jj}), the components of the Killing spinor that do depend
on the compact coordinate have to be projected out of the
dimensionally reduced theory \cite{Bergshoeff:1994cb}, leading to less
supersymmetry.  Since
T~duality of classical solutions involves their dimensional reduction
it should not come as a surprise that the supersymmetry of the
maximally supersymmetric vacua is not preserved by T~duality either
\cite{Bakas:1994ba,Duff:1997qz}.

In this paper we are going to show that the known maximally supersymmetric 
$d=4,5,6$ vacua of theories with 8 supercharges
($N=2$ or $N=(2,0)$ theories) can be dimensionally reduced/oxidized
preserving all their unbroken supersymmetries because in all the
$d=5,6$ cases it is possible to choose coordinates in which the
Killing spinor is independent of the coordinate we use for
dimensional reduction.

That the coordinate choice that preserves all supersymmetry in
dimensional reduction is always possible looks highly non-trivial.
However, thinking in terms of oxidation of the $d=4,5$ theories it is
evident that all unbroken supersymmetry should be preserved: these
theories can be obtained by standard dimensional reduction of the
$d=6,5$ ones supplemented by a truncation of the matter multiplets
that appear in the reduction. It is, therefore, guaranteed that, if we
have a solution of the $d=4,5$ theories that preserves all 8
supersymmetries, it comes from some $d=5,6$ solution that also
preserves those 8 supersymmetries and therefore has to be one of the
essentially unique maximally supersymmetric vacua of the
theory~\footnote{To the best of our knowledge, though, no theorem
  proving the uniqueness of the maximally supersymmetric
  vacua we are dealing with exists for the $d=6$, $N=(2,0)$ theory. A
  classification of the spacetimes admitting Killing spinors in four
  dimensions was given in \cite{Tod:pm}, and a complete classification
  of supersymmetric solutions in $d=5$, $N=2$ supergravity has recently
  appeared \cite{Gauntlett:2002nw}.}. Thus, the maximally
supersymmetric vacua of these theories must be related by dimensional
reduction/oxidation and we are going to show exactly how this happens.
The independence of the Killing spinors of the compact coordinates is
an implicit automatic consequence of the above arguments.
\par
Let us now briefly review the known maximally supersymmetric vacua of
these theories:

\begin{description}
  
\item[$N=(2,0), d=6$:]  \hfill
  \begin{enumerate}
  \item The 1-parameter family of Kowalski-Glikman (KG) H$pp$-wave
    solutions found in Ref.~\cite{Meessen:2001vx} that we will denote
    by $KG6(2,0)$.
  \item The 1-parameter family of solutions with $aDS_{3} \times
    S^{3}$ geometry found in Ref.~\cite{Gibbons:1994vm} as the
    near-horizon limit of the self-dual string solution.
\end{enumerate}
  
\item[$N=2, d=5$:]  \hfill
  \begin{enumerate}
  \item The 1-parameter family of KG solutions solutions found in
    Ref.~\cite{Meessen:2001vx} that we will denote by $KG5$. 
  \item The 1-parameter family of solutions with $aDS_{3} \times
    S^{2}$ geometry found in Ref.~\cite{Gibbons:1994vm} as
    near-horizon limit of the extreme string solution.
  \item  The 1-parameter family of solutions with $aDS_{2}
    \times S^{3}$ geometry found in Ref.~ \cite{Kallosh:1996vy} as
    near-horizon limit of the extreme black hole solution.
  \item The 2-parameter family of $N=2, d=5$ solutions found in
    Ref.~\cite{Gauntlett:1998fz} as the near-horizon limit of the
    supersymmetric rotating black hole solution.
  \end{enumerate}

The third family is contained in the fourth and corresponds to a
vanishing rotation parameter. We will show that the second family is
also contained in the fourth and corresponds to the value $1$ of the
rotation parameter.

\item[$N=2, d=4$:] \hfill 
  \begin{enumerate}
  \item The 1-parameter\footnote{Electric-magnetic duality rotations
      only change the polarization plane of an electromagnetic wave
      and their effect on this family of solutions can be undone by a
      rotation that leaves the form of the metric invariant.} family of KG
    solutions solutions found in Ref.~\cite{Kowalski-Glikman:1985im}
    that we will denote by $KG4$.
    
  \item The 2-parameter family of electric/magnetic $N=2,d=4$
    Robinson-Bertotti solutions \cite{kn:Rob} that have the
    geometry $aDS_{2} \times S^{2}$.
  \end{enumerate}

\end{description}

\begin{figure}[!ht]
\begin{center}
\leavevmode
\epsfxsize= 10cm
\epsffile{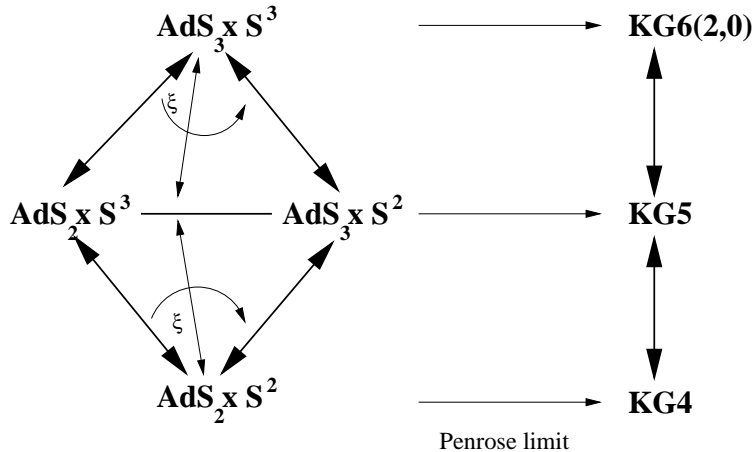}
\caption{Relations between the $d=4,5,6$ vacua with 8 supercharges.}
\label{fig:d456vacua}
\end{center}
\end{figure}

The connections between these vacua that we have found are summarized
in Figure~\ref{fig:d456vacua}\footnote{The left hand side of the
  relations of the relations were discussed in \cite{Boonstra:1998yu}.
  We thank K.~Skenderis for pointing this out to us.}. The relations
between the $KG$ solutions are straightforward. The $aDS_{3}\times
S^{3}$ can be dimensionally reduced in the direction of the $S^{1}$
Hopf fiber of the 3-sphere and then we get $aDS_{3}\times S^{2}$. It
can also be reduced in the $S^{1}$ fiber of the $aDS_{3}$, giving
$aDS_{2}\times S^{3}$. Finally, we can rotate these fibers an angle
$\xi$ and reduce, getting the maximally supersymmetric solution that
is the near-horizon limit of the rotating 5-dimensional extreme black
hole. The angular momentum parameter $j$ is essentially $\sin \xi$.
Thus, this 2-parameter family of 5-dimensional vacua interpolates (in
parameter space) between the $aDS_{2}\times S^{3}$ and the
$aDS_{3}\times S^{2}$ vacua. The reduction of any member of this
family in the remaining fiber gives an electric/magnetic
Robinson-Bertotti solution where $\sin \xi$ is the ratio between
the electric and the magnetic fields.

This paper is organized as follows: in Section~\ref{sec-Hpp} we study
how the $KG6(2,0)$ and $KG5$ solutions can be dimensionally reduced
preserving all the supersymmetry after describing briefly the general
form of $pp$-wave solutions and their sources in
Section~\ref{sec-generalppwavesolutions}.  In Section~\ref{sec-AdSS}
we study how the $aDS_{m}\times S^{n}$-type vacua of these theories
are related by oxidizing them. 
Section~\ref{sec-conclusions} contains our conclusions and some discussion.
\section{Dimensional Reduction of Maximally Supersymmetric $Hpp$-Waves}
\label{sec-Hpp}
Before we study the reduction of $KG$ solutions it is worth studying
briefly general supergravity $pp$-wave solutions.
\subsection{General $pp$-Wave Solutions}
\label{sec-generalppwavesolutions} 
$pp$-waves spacetimes are those whose metric admits a covariantly
constant null vector. A metric with this property can always be put in
the form 

\begin{equation}
ds^{2} = 2du(dv+Kdu+{\cal A}_{a}dx^{a}) +\tilde{g}_{ab}dx^{a}dx^{b}\, ,
\end{equation}

\noindent where the functions $K,{\cal A}_{a},\tilde{g}_{ab}$ depend
only on the wave-front coordinates $x^{a}$ and on the null coordinate
$u$. ${\cal A}_{a}$ is known as the \textit{Sagnac connection}
\cite{Gibbons:1999uv} and can always be set to zero by means of a
coordinate transformation.

In supergravity theories it is natural to look for $pp$-wave solutions
of the system 

\begin{equation}
  S_{a}={\textstyle\frac{1}{16\pi G_{N}^{(d)}}}
    \int{d^{d}x\sqrt{|g|}\left[R+{\textstyle\frac{1}{2}}(\partial\varphi)^2+
    \textstyle{\frac{(-1)^{p+1}}{2(p+2)!}}
    e^{-2a\varphi}F_{(p+2)}^2\right]}\, , 
\end{equation}

\noindent where $F_{(p+2)}=dA_{(p+1)}$, of the form

\begin{equation}
\left\{
  \begin{array}{rcl}
  ds^{2} & = & 2du(dv+Kdu)+\tilde{g}_{ab}dx^{a}dx^{b}\, , \\ \\ 
  F_{(p+2)} & = &  du\land C\, ,  \\
  \end{array}
\right.
\end{equation}

\noindent where $C$ is a $(p+1)$-form on the wave-front space and,
as all the other fields in this Ansatz, it is independent of $v$.

A general solution is provided by a Ricci-flat wave-front metric
$\tilde{g}_{ab}$ 
which must also satisfy

\begin{displaymath}
  \tilde{\nabla}_a\left(\tilde{g}^{bc}\partial_u\tilde{g}_{bc}\right)-
  \tilde{\nabla}_b\left(\tilde{g}^{bc}\partial_u\tilde{g}_{ac}\right)=0\, ,    
\end{displaymath}

\noindent a harmonic $(p+1)$-form $C$ in wave-front space

\begin{displaymath}
  \tilde{d}C=\tilde{d}^{\tilde{\star}}C=0\, , 
\end{displaymath}

\noindent with arbitrary $u$-dependence, an arbitrary function 
$\varphi(u)$; and a function $K(u,x^a)$ satisfying the equation

\begin{displaymath}
  \tilde{\nabla}^2K+
  {\textstyle\frac{1}{4}}\partial_u\tilde{g}^{ab}\partial_u\tilde{g}_{ab}+
  {\textstyle\frac{1}{2}}\tilde{g}^{ab}\partial_u^2\tilde{g}_{ab}+
  {\textstyle\frac{1}{2}}(\partial_u\varphi)^2+
  {\textstyle\frac{(-1)^{p+1}}{2(p+1)!}}e^{-2a\varphi}C^2=0\, .  
\end{displaymath}

The simplest choice of Ricci-flat wave-front space leads to the solutions

\begin{equation}
  \label{Simplest}
  \begin{array}{rcl}
\tilde{g}_{ab} & = & -\delta_{ab}\, , 
\hspace{1cm}
C= C(u)\, ,
\hspace{1cm}
 \varphi = \varphi(u)\, , \\
& & \\  
K & = & H +A\, ,
\hspace{1cm}
A\equiv A_{ab}(u) x^{a}x^{b}=  -{\textstyle\frac{1}{4}}
\left[(\partial_u\varphi)^2+
{\textstyle\frac{(-1)^{p+1}}{(p+1)!}}e^{-2a\varphi}C^2\right]
(\mbox{tr}M)^{-1}M_{ab}\, x^{a}x^{b}\, ,  
\end{array}
\end{equation}

\noindent where $H= H(x^a)$ is an arbitrary harmonic function 
in wave-front space, $M_{ab}$ is a constant symmetric matrix and $C$
and $\varphi$ are just arbitrary functions of $u$.

The function $K$ that contains all the information has, therefore,
two pieces: the harmonic function $H(x)$, independent of the gauge
field and dilaton (i.e.~purely gravitational), and the matrix
$A_{ab}(u)$ that depends on the gauge field and dilaton. One can argue
that $H$ represents excitations over a vacuum that consists of a
self-supported (source-less) gauge field and dilaton and a metric
described by $A_{ab}(u)$. For instance, one can try to match the above
solution with a charged, mass-less, $p$-brane source with effective
action

\begin{equation}
\begin{array}{rcl}
  S_{p}[X^{\mu},\gamma_{ij}] & = & 
  -T_p{\displaystyle\int} d^{p+1}\xi\sqrt{|\gamma|}\,
  e^{-2b\varphi}\, \gamma^{ij}\partial_{i}X^{\mu} \partial_{j}X^{\nu} 
  g_{\mu\nu} + \\
  & & \\
  & & 
+\frac{(-1)^{p+1}\mu_p}{(p+1)!}{\displaystyle\int} 
d^{p+1}\xi\, A_{(p+1)\, \mu_{1}\ldots\mu_{p+1}}
\partial_{i_{1}}X^{\mu_{1}}\ldots\partial_{i_{p+1}}X^{\mu_{p+1}}
\epsilon^{i_1\ldots i_{p+1}}\, .
\end{array}
\end{equation}

The following Ansatz\footnote{Such an Ansatz was also discussed in
Ref. \cite{Skenderis:2002vf} for probing the type IIB's KG wave, but was
found to be inconsistent. Here such trouble is avoided because,
contrary to Ref. \cite{Skenderis:2002vf}, a massless $p$-brane is used.}

\begin{equation}
  U(\xi)=0\, , \hspace{1cm} V(\xi) = \alpha\, \xi^{0}\, , \hspace{1cm} 
  X^a(\xi)= 0\, , \hspace{1cm} \sqrt{|\gamma|}\, \gamma^{00}=1\, ,
\end{equation}

\noindent $\alpha$ being some constant and $\xi^{0}$ being the worldvolume
time coordinate (plus the above values Eq.~(\ref{Simplest}) for the
spacetime fields) representing the brane moving in a direction
transverse to its worldvolume reduces all the equations of motion to
only one which is not automatically satisfied,\footnote{For $p=0$ the
  charge $\mu_{0}$ has to be set to zero in order to satisfy the
  equation of motion for that gauge field, but in the other cases the
  value of $\mu_{p}$ does not play any role.} {\em i.e.}

\begin{equation}
  \partial_{a}\partial_{a} H=
  16\pi G_N^{(d)}T_{p} 
  e^{-2b\varphi(0)}\alpha^{2}\delta(u)\delta(\vec{x}_8)\, .
\end{equation}

Thus, only $H$ feels the source and the gauge field seems to be
self-supported. The solutions with $H=0$ can be interpreted as vacua
and can be described as homogeneous spaces
\cite{kn:CaWa,Figueroa-O'Farrill:2001nz} (H$pp$-waves).  Actually, the
presence of a covariantly constant null vector ensures that at least
half of the supersymmetries will always be unbroken if we embed the
above solutions in a supergravity theory (even for $H\neq 0$) but in
some cases (the Kowalski-Glikman solutions
\cite{Kowalski-Glikman:wv,Blau:2001ne,Meessen:2001vx,Kowalski-Glikman:1985im}) 
there are H$pp$-wave solutions that preserve all the supersymmetries.
See \cite{Papadopoulos:1999tg} for a discussion
on waves that preserve fractions of the supersymmetry.
\par
It was recently shown \cite{Michelson:2002wa} that the maximal amount of 
supersymmetry that can be preserved in a circle compactification of 
the KG10 solution \cite{Blau:2001ne} is $3/4$ and the same thing holds
for the 11-dimensional KG wave \cite{Kowalski-Glikman:wv}.
Although one would expect the same to happen in the $N=2$ $d=6,5,4$
KG-solutions, we are going to show that they are related by
dimensional reduction.  First of all, the susy preserving 
dimensional reduction is
possible after a change of coordinates in which the dependence on the
compact coordinate is removed at the expense of introducing a
non-vanishing Sagnac connection. It turns out that in the new
coordinates the Killing spinors 
are independent of the compact coordinates so that dimensional
reduction will preserve all of them. Furthermore, the Sagnac connection
becomes a KK vector that combines in the right way with the other vector
fields present to cancel the matter multiples that arise in the two
dimensional reductions involved.
\subsection{Reduction of $KG6(2,0)$ to $KG5$}
$N=(2,0),d=6$ supergravity\footnote{Our conventions are essentially
  those of Ref.~\cite{Nishino:dc} with some changes in the
  normalizations of the fields. In particular
  $\gamma_{7}=\gamma_{0}\cdots \gamma_{5}$, $\gamma_{7}^{2}=+1$,
  $\epsilon^{012345}=+1$, $\gamma^{a_{1}\cdots a_{n}} =
  \frac{(-1)^{[n/2]}}{(6-n)!}  \epsilon^{a_{1}\cdots a_{n}b_{1}\cdots
    b_{6-n}}\gamma_{b_{1}\cdots b_{6-n}} \gamma_{7}$. Positive and
  negative chiralities are defined by $\gamma_{7}\hat{\psi}^{\pm}=\pm
  \hat{\psi}^{\pm}$.} consists of the metric
$\hat{e}^{\hat{a}}{}_{\hat{\mu}}$, 2-form field
$\hat{B}^{-}_{\hat{\mu}\hat{\nu}}$ with anti-self-dual field strength
$\hat{H}^{-}=3\partial\hat{B}^{-}$ and positive-chirality symplectic
Majorana-Weyl gravitino $\hat{\psi}^{+}_{\hat{\mu}}$. The bosonic
equations of motion can be derived from the action

\begin{equation}
\hat{S}=\int d^{6}\hat{x}\sqrt{|\hat{g}|}\, [\hat{R}
+{\textstyle\frac{1}{12}} \hat{H}^{2}]\, ,  
\end{equation}

\noindent imposing afterwards the anti-self-duality constraint 
${}^{\star}\hat{H}^{-}=-\hat{H}^{-}$. The gravitino supersymmetry
transformation rule is (for zero fermions)

\begin{equation}
\delta_{\hat{\epsilon}^{+}}\hat{\psi}^{+}_{\hat{a}} 
=\left(\hat{\nabla}_{\hat{a}}  
-{\textstyle\frac{1}{48}}\not\!\!\hat{H}^{-}\hat{\gamma}_{\hat{a}} 
\right)\hat{\epsilon}^{+}\, .
\end{equation}

This can be reduced to $N=2,d=5$ supergravity (metric $e^{a}{}_{\mu}$,
{\it graviphoton} vector field $\mathcal{V}_{\mu}$ and
symplectic-Majorana gravitino $\psi_{\mu}$) coupled to a vector
multiplet consisting of a gaugino (the 6th component of the
6-dimensional gravitino, a real scalar (the KK one) and a vector field
$\mathcal{W}_{\mu}$. The vector fields $\mathcal{V}_{\mu}$ and
$\mathcal{W}_{\mu}$ are combinations of scalars, the KK vector field
that comes from the 6-dimensional metric $A_{\mu}$ and the vector
field that comes from the 6-dimensional 2-form $B_{\mu}$. The
identification of the right combinations will be made by imposing
consistency of the truncation.

Using the same techniques as in the reduction of $N=2B,d=10$
supergravity on a circle Ref.~\cite{Meessen:1998qm} one gets the
5-dimensional action

\begin{equation}
\label{eq:N=2d=5action}
S=\int d^{5}x\sqrt{|g|}\, k\,[R -{\textstyle\frac{1}{4}}k^{2}F^{2}(A)
-{\textstyle\frac{1}{4}}k^{-2}F^{2}(B)
+{\textstyle\frac{\epsilon}{8\sqrt{|g|}}}k^{-1}F(A)F(B)B]\, .   
\end{equation}

The truncation to pure supergravity involves setting $k=1$
consistently, i.e.~in such a way that its equation of motion is always
satisfied. The $k$ equation of motion with $k=1$ (upon use of
Einstein's equation) implies the constraint 

\begin{equation}
F^{2}(B)=2F^{2}(A)\, .  
\end{equation}

Let us introduce two linear combinations
$\mathcal{F}(\mathcal{V}),\mathcal{G}(\mathcal{W})$ of the vector
field strengths

\begin{equation}
\left\{
  \begin{array}{rcl}
\mathcal{F}(\mathcal{V}) & = & \alpha F(A) +\beta F(B)\, ,\\
& & \\
\mathcal{G}(\mathcal{W}) & = & -\beta F(A) +\alpha F(B)\, ,\\
\end{array}
\right.
\end{equation}

\noindent with $\alpha^{2}+\beta^{2}=1$. Substituting them into the 
above constraint we see that it is automatically satisfied with
$\mathcal{G}(\mathcal{W})=0$ and $\beta^{2}=2\alpha^{2}$, so $\alpha
=s(\alpha)/\sqrt{3}$ and $\beta =s(\beta)\sqrt{2}/\sqrt{3}$.
These conditions reduce the equations of motion of $A$ and $B$ to a single
equation for $\mathcal{V}$. This equation and the resulting Einstein
equation can be derived from the action

\begin{equation}
S=\int d^{5}x\sqrt{|g|}\,[R
-{\textstyle\frac{1}{4}}\mathcal{F}^{2}
+s(\alpha)
{\textstyle\frac{\epsilon}{12\sqrt{3} \sqrt{|g|}}}
\mathcal{F}\mathcal{F}\mathcal{V}]\, , 
\end{equation}

\noindent which is that of the bosonic sector of $N=2,d=5$ supergravity 
\cite{Cremmer:1980gs}. The relative sign of $\alpha$ and $\beta$ will
be fixed by supersymmetry: using the decomposition

\begin{equation}
  \begin{array}{rcl}
\hat{\gamma}^{a} & = & \gamma^{a}\otimes \sigma^{1}\, ,
\hspace{1cm} a=0,1,2,3,4\, ,\\
& & \\
\hat{\gamma}^{5} & = & \mathbb{I}\otimes i\sigma^{2}\, ,\\
& & \\
\hat{\gamma}_{7} & = & \hat{\gamma_{0}}\cdots\hat{\gamma}_{5}
=\mathbb{I}\otimes \sigma^{3}\, ,\\
  \end{array}
\end{equation}

\noindent where the $\gamma^{a}$s are 5-dimensional gamma matrices 
satisfying $\gamma_{0}\cdots\gamma_{4}=\mathbb{I}$, using chirality,
we can split the gravitino supersymmetry transformation rule into

\begin{equation}
\left\{
  \begin{array}{rcl}
\delta_{\epsilon}\hat{\psi}_{a} & = & \left\{ \nabla_{a}
-\frac{1}{8\sqrt{2}}k ^{-1}\not\!\!F(B)\gamma_{a} 
-\frac{1}{4}k \not\!F_{a}(A)\right\}\epsilon\, ,\\
& & \\
\delta_{\epsilon}\hat{\psi}_{w} & = & \left\{\partial_{w}
+\frac{1}{2}\not\! \partial \log{k} +\frac{1}{8}k \not\!\!F(A) 
-\frac{1}{8\sqrt{2}}k^{-1} \not\!\!F(B)\right\}\epsilon\, .\\
  \end{array}
\right.
\end{equation}

\noindent $\hat{\psi}_{w}$ is the 5-dimensional gaugino and its 
supersymmetry transformation has to be identically zero. This can be
achieved by taking $\epsilon$ independent of $w$ and identifying
$s(\alpha)=s(\beta)$ so

\begin{equation}
\mathcal{G} \equiv s(\alpha) 
\left({\textstyle\frac{1}{\sqrt{3}}}F(B)
-\sqrt{\textstyle\frac{2}{3}}F(A)\right)\equiv 0\, . 
\end{equation}

It only remains the supersymmetry transformation law of
$\hat{\psi}_{a}$ that becomes the 5-dimensional gravitino. Expressed
in terms of the surviving vector field, it takes the right
form\footnote{Actually, either the sign of the Chern-Simons term or
  the $\mathcal{F}$ term in the supersymmetry transformation rule in
  Ref.~\cite{Cremmer:1980gs} is wrong. Choosing the sign of $\alpha$
  we can make either of them coincide with those in
  Eqs.~(\ref{eq:N=2d=5action}) and (\ref{eq:N=2d=5susyrule}), but
  not both at the same time. A further check of these signs is
  provided by the reduction to $d=4$: the consistency conditions for
  the truncation to pure $N=2,d=4$ supergravity coming from the action
  and the gaugino supersymmetry transformation rule are incompatible
  with the signs of Ref.~\cite{Cremmer:1980gs} but fully compatible
  with ours.}  \cite{Cremmer:1980gs}

\begin{equation}
\label{eq:N=2d=5susyrule}
\delta_{\epsilon}\psi_{a}=
\left\{ \nabla_{a}
-s(\alpha){\textstyle\frac{1}{8\sqrt{3}}}
(\gamma^{bc}\gamma_{a}+2\gamma^{b}g^{c}{}_{a})
\mathcal{F}_{bc}\right\}\epsilon\, .
\end{equation}

The relation between 6- and 5-dimensional pure supergravity fields is

\begin{equation}
  \begin{array}{|rcl|rcl}
\hat{g}_{\underline{w}\underline{w}} & = & -1\, , &
\hat{B}^{-}_{\mu\underline{w}} & = &
\frac{s(\alpha)}{\sqrt{3}}\mathcal{V}_{\mu}\, ,\\
& & & & & \\
\hat{g}_{\mu\underline{w}} & = & 
\frac{s(\alpha)}{\sqrt{3}}\mathcal{V}_{\mu}\, , \hspace{.7cm} &
\hat{g}_{\mu\nu} & = & g_{\mu\nu} -\frac{1}{3}
\mathcal{V}_{\mu}\mathcal{V}_{\nu}\, ,\\
  \end{array}
\end{equation}

\noindent while the $\hat{B}^{-}_{\mu\nu}$ components can be found
imposing anti-self-duality.

Now, let us consider the $KG6(2,0)$ solution \cite{Meessen:2001vx} in
canonical coordinates with $\hat{B}^{-}$ in a convenient gauge

\begin{equation}
\label{eq:KG6canonical}
KG6(2,0): \,\,\,\,
\left\{
  \begin{array}{rcl}
d\hat{s}^{2} & = & 2du[dv +\frac{\lambda_{6}^{2}}{8}\vec{x}_{(4)}^{\, 2}du]
-d\vec{x}_{(4)}^{2}\, ,
\hspace{1cm}\vec{x}_{(4)}\equiv (x,y,z,w)\, ,\\
& & \\
\hat{B}^{-} & = & \lambda_{6}du\land (zdw-xdy)\, .\\
  & & \\
 \hat{\epsilon} &=& \left[
                        1-\frac{\lambda_{6}}{4}\hat{\gamma}^{+23}
                          \vec{x}_{(4)}\cdot \hat{\vec{\gamma}}
                    \right]
                 \exp\left(
                        \frac{u\lambda_{6}}{4}
                            \hat{\gamma}^{+23}\hat{\gamma}^{-}
                     \right)\ \hat{\epsilon}^{(0)} \; ,
  \end{array}
\right.
\end{equation}

Performing the coordinate transformations

\begin{equation}
\label{eq:KG6_coordinate_trafo}
\left\{
  \begin{array}{rcl}
z & = & \cos{(\frac{\lambda_{6}}{2}u)}z^{\prime} 
+\sin{(\frac{\lambda_{6}}{2}u)}w^{\prime}\, ,\\
& & \\
w & = & -\sin{(\frac{\lambda_{6}}{2}u)}z^{\prime} 
+\cos{(\frac{\lambda_{6}}{2}u)}w^{\prime}\, ,\\
& & \\ 
v & = & v^{\prime} +\frac{\lambda_{6}}{2}z^{\prime}w^{\prime}\, ,\\
  \end{array}
\right.
\end{equation}

\noindent the solution takes the $w^{\prime}$-independent form

\begin{equation}
\label{eq:KG6w-independent}
KG6(2,0): \,\,\,\,
\left\{
  \begin{array}{rcl}
\hat{s}^{2} & = & 2du[dv^{\prime} +\frac{\lambda_{6}^{2}}{8}(x^{2}+y^{2})du
+\lambda_{6}z^{\prime}dw^{\prime}]
-d\vec{x}_{(4)}^{\, \prime\,2}\, ,
\hspace{1cm}\vec{x}_{(4)}^{\prime}\equiv (x,y,z^{\prime},w^{\prime})\, ,\\
& & \\
\hat{B}^{-} & = & \lambda_{6}du\land(z^{\prime}dw^{\prime}-xdy)\, .\\
 & & \\
 \hat{\epsilon} &=& \left[
                        1+\frac{\lambda_{6}}{4}\hat{\gamma}^{+23}
                        \left\{
                          x\hat{\gamma}_{2} +
                          y\hat{\gamma}_{3}
                        \right\}
                    \right]
                    \exp\left(
                        \frac{u\lambda_{6}}{4}
                        \left\{ 
                            \hat{\gamma}^{45}+               
                            \hat{\gamma}^{+23}\hat{\gamma}^{-}
                        \right\}
                    \right)\ \hat{\epsilon}^{(0)} \; .
  \end{array}
\right.
\end{equation}

It is easy to see that it satisfies the truncation conditions

\begin{equation}
\hat{g}_{\underline{w}\underline{w}}=-1\, ,
\hspace{.5cm}
\hat{B}^{-}_{\mu\underline{w}}=\hat{g}_{\mu\underline{w}}\, , 
\hspace{.5cm}
\partial_{\underline{w}}\hat{\epsilon}^{+}=0\, ,
\end{equation}

\noindent and, thus, it can be reduced to a solution of pure $N=2,d=5$ 
supergravity that turns out to be the maximally supersymmetric $KG5$
solution \cite{Meessen:1998qm}:

\begin{equation}
\label{eq:KG5canonical}
KG5: \,\,\,\,
\left\{
  \begin{array}{rcl}
ds^{2} & = & 2du[dv^{\prime} 
+\frac{\lambda_{5}^{2}}{24}(4z^{\prime\, 2}+ x^{2}+y^{2})du]
-d\vec{x}_{(3)}^{\, \prime\,2}\, ,
\hspace{1cm}\vec{x}_{(3)}^{\, \prime}\equiv (x,y,z^{\prime})\, ,\\
& & \\
\mathcal{F} & = & \lambda_{5}du\land dz^{\prime}\, ,
\hspace{1cm}
\lambda_{5}=-s(\alpha)\sqrt{3}\lambda_{6}\, .\\
  \end{array}
\right.
\end{equation}


\subsection{Reduction of $KG5$ to $KG4$}

The action Eq.~(\ref{eq:N=2d=5action}) can be straightforwardly
reduced to $d=4$ dimensions giving the action of $N=2,d=4$
supergravity (consisting of the metric, the {\it graviphoton} vector
field $V_{\mu}$ and a gravitino) coupled to a vector multiplet
(consisting of a vector $W_{\mu}$ and two real scalars $k,l$ plus a
gaugino) \cite{Chamseddine:1980sp}. The two vectors will be
combinations of the KK vector $A_{\mu}$ that comes from the metric and
the vector $B_{\mu}$ that comes from the 5-dimensional vector
$\mathcal{V}_{\mu}$. To determine the right combinations, we study the
consistency of the truncation of the fields that belong for sure to
the matter multiplet $k=1,l=0$ and the gaugino.

The action for the 4-dimensional bosonic fields is 

\begin{equation}
  \begin{array}{rcl}
S & = &{\displaystyle\int} d^{4}x\sqrt{|g|}\, k\, \left\{R 
+{\textstyle\frac{1}{2}}k^{-2}(\partial l)^{2}
-{\textstyle\frac{1}{4}} k^{2}F^{2}(A) 
-{\textstyle\frac{1}{4}} [F(B)+lF(A)]^{2} \right.\\
& & \\
& & 
\hspace{1cm}
\left.
+s(\alpha)\frac{k^{-1}l}{4\sqrt{3}\sqrt{|g|}}
\epsilon[F(B)+lF(A)-2A\partial l]^{2}\right\}\, .\\
\end{array}
\end{equation}

Setting $k=1,l=0$ in the equations of motion of $k$ and $l$ we get two
constraints:

\begin{equation}
\left\{
  \begin{array}{rcl}
3F^{2}(A) +F^{2}(B) & = & 0\, ,\\
& & \\
\sqrt{3}F(A)-s(\alpha){}^{\star}F(B) & = & 0\, .\\
  \end{array}
\right.
\end{equation}

The second constraint implies the first and is actually sufficient to
identify the graviphoton and the matter vector field
strengths\footnote{For $k=1,l=0$ only.} up to global, irrelevant,
signs, that we fix arbitrarily

\begin{equation}
\left\{
  \begin{array}{rcl}
F(V) & = & \frac{1}{2}{}^{\star} F(A) -s(\alpha)\frac{\sqrt{3}}{2}F(B)\, ,\\
& & \\
F(W) & = & -\frac{\sqrt{3}}{2}{}^{\star} F(A)-s(\alpha)\frac{1}{2}F(B)\, .\\
  \end{array}
\right.  
\end{equation}

Setting $F(W)=0$ (which is consistent with the $W_{\mu}$ equation of
motion) we get the action of (the bosonic sector of) pure $N=2,d=4$
supergravity (the Einstein-Maxwell action)

\begin{equation}
S = \int d^{4}x\sqrt{|g|}\, [R -{\textstyle\frac{1}{4}} F^{2}(V)]\, . 
\end{equation}

We can see that this truncation is consistent with the supersymmetry
transformation rules. The 5-dimensional matrices
$\hat{\gamma}^{\hat{a}}$ decompose into 4-dimensional matrices as follows:

\begin{equation}
\hat{\gamma}^{a}=\gamma^{a}\, ,\,\,\,\, a=0,1,2,3\, ,
\hat{\gamma}^{4}=-i \gamma_{5}=\gamma_{0}\gamma_{1}\gamma_{2}\gamma_{3}\, .  
\end{equation}

The 5-dimensional symplectic-Majorana spinors are a pair of ordinary
4-component Dirac spinors related by the symplectic-Majorana
constraint. Thus, in $d=4$ we simply keep one of them, which will be
unconstrained and decomposable, if necessary, into a pair of
4-dimensional Majorana spinors.

Now if the supersymmetry parameter is independent of
the compactification direction $y$ and we set $k=1,l=0$ in the $y$
component of the gravitino transformation rule (which should become
the gaugino transformation rule), we find that

\begin{equation}
\delta_{\epsilon}\hat{\psi}_{y}=
{\textstyle\frac{i}{4\sqrt{3}}}\not\!\!F(W)\epsilon\, .  
\end{equation}

\noindent and, so, the truncation $F(W)=0$ is consistent with setting the 
gaugino to zero. The supersymmetry transformation rule of the surviving
gravitino is

\begin{equation}
\delta_{\epsilon}\psi_{a} =\left[\nabla_{a}+{\textstyle\frac{1}{8}}
\not\!\!F(V)\gamma_{a} \right]\epsilon \, . 
\end{equation}

The relation between 5-dimensional and 4-dimensional fields that satisfy the
truncation condition is

\begin{equation}
  \begin{array}{|rcl|rcl}
\hat{g}_{\underline{y}\underline{y}} & = & -1\, , &
\hat{\mathcal{V}}_{\underline{y}} & = & 0\, ,\\
& & & & & \\
2\partial_{[\mu}\hat{g}_{\nu]\underline{y}} & = & 
\frac{-1}{4\sqrt{|g|}}\epsilon_{\mu\nu\rho\sigma}F^{\rho\sigma}(V)\, , 
\hspace{1cm}&
\hat{\mathcal{V}}_{\mu} & = & -s(\alpha)\frac{\sqrt{3}}{2} V_{\mu}\, ,\\
& & & & & \\
\hat{g}_{\mu\nu} & = & g_{\mu\nu}
-\hat{g}_{\mu\underline{y}}\hat{g}_{\nu\underline{y}}\, ,& & & \\
  \end{array}
\end{equation}

Now, to apply these results to the $KG5$ solution Eq.~(\ref{eq:KG5canonical})
we first perform the change of coordinates

\begin{equation}
\label{eq:KG5_coordinate_trafo}
\left\{
  \begin{array}{rcl}
x & = & \cos{(\frac{\lambda_{5}}{2\sqrt{3}}u)}x^{\prime} 
+\sin{(\frac{\lambda_{5}}{2\sqrt{3}}u)}y^{\prime}\, ,\\
& & \\
y & = & -\sin{(\frac{\lambda_{5}}{2\sqrt{3}}u)}x^{\prime} 
+\cos{(\frac{\lambda_{5}}{2\sqrt{3}}u)}y^{\prime}\, ,\\
& & \\ 
v^{\prime} & = & v^{\prime\prime} 
+\frac{\lambda_{5}}{2\sqrt{3}}x^{\prime}y^{\prime}\, ,\\
  \end{array}
\right.
\end{equation}

\noindent that puts the $KG5$ solution in the $y^{\prime}$-independent form

\begin{equation}
\label{eq:KG5y-independent}
KG5: \,\,\,\,
\left\{
  \begin{array}{rcl}
ds^{2} & = & 2du[dv^{\prime\prime} 
+\frac{\lambda_{5}^{2}}{6}z^{\prime\, 2}du
+\frac{\lambda_{5}}{\sqrt{3}}x^{\prime}dy^{\prime}
]
-d\vec{x}_{(3)}^{\, \prime\,2}\, ,
\hspace{1cm}\vec{x}_{(3)}^{\, \prime}\equiv (x^{\prime},y^{\prime},z^{\prime})\, ,\\
& & \\
\mathcal{F} & = & \lambda_{5}du\land dz^{\prime}\, .
  \end{array}
\right.
\end{equation}

\noindent In this form, the $KG5$ solution just happens to 
satisfy the truncation condition that allows us to reduce it to a pure
$N=2,d=4$ supergravity solutions that turns out to be the $KG4$
maximally supersymmetric spacetime \cite{Kowalski-Glikman:1985im}, as
promised

\begin{equation}
\label{eq:KG4canonical}
KG4: \,\,\,\,
\left\{
  \begin{array}{rcl}
ds^{2} & = & 2du[dv^{\prime\prime} 
+\frac{\lambda_{4}^{2}}{8}\vec{x}_{(2)}^{\prime}{}^{2}\ du]
-d\vec{x}_{(2)}^{\, \prime\,2}\, ,
\hspace{1cm}\vec{x}_{(2)}^{\, \prime}\equiv (x^{\prime},z^{\prime})\, ,\\
& & \\
F & = & \lambda_{4}du\land dz^{\prime}\, ,
\hspace{1cm}
\lambda_{4}=s(\alpha)\frac{2}{\sqrt{3}}\lambda_{5}\, .\\
  \end{array}
\right.
\end{equation}

At first sight it is surprising that in all cases the truncation
condition can be satisfied, at least in a certain gauge.  Actually, it
is easy to see that it must happen by thinking in terms of oxidation
of the lower-dimensional solutions: Since the $N=2,d=5$ theory can be
reduced to $N=2,d=4$ supergravity coupled to a vector multiplet that
can be consistently truncated, any solution of pure $N=2,d=4$
supergravity can be uplifted to a solution of the $N=2,d=5$ theory
with the same, or bigger, amount of supersymmetry.
Therefore, the $KG4$ solution
can be uplifted to a maximally supersymmetric solution of the $N=2,d=5$
theory which turns out to be the $KG5$ solution in non-canonical
coordinates.  Essentially the same mechanism works in the oxidation of
the $KG5$ solution to a maximally supersymmetric solution of
$N=(2,0),d=6$ that turns out to be the $KG6(2,0)$.

Now it is clear that the same should happen in all cases: all
solutions of pure $N=2,d=4$ supergravity must be related via
dimensional reduction/oxidation to pure $N=2,d=5$ and $N=(2,0),d=6$
supergravity solutions that preserve the same amount of supersymmetry.
In particular, maximally supersymmetric solutions of these three
theories should be related. We have seen that this is true for the
$KG$ spacetimes and now we are going to study the $aDS_{n}\times
S^{m}$ spacetimes.

\section{Oxidation of Maximally Supersymmetric $d=4,5,6$ 
  $aDS_{n}\times S^{m}$ Spacetimes}
\label{sec-AdSS}
\subsection{Oxidation of the Robinson-Bertotti Solution}
The Robinson-Bertotti solution \cite{kn:Rob} can be obtained
either as a particular member of the Majumdar-Papapetrou family of
solutions of the Einstein-Maxwell equations \cite{kn:MaPa} or as
the near-horizon limit of the extreme Reissner-Nordstr{\"o}m black hole
solution \cite{Gibbons:1984kp} and is given in its electric and magnetic
versions by

\begin{equation}
\label{eq:ads2xs2}
\left\{
  \begin{array}{rcl}
ds^{2} & = & R_{2}^{2}\,  d\Pi_{(2)}^{2} 
-R_{2}^{2}\, d\Omega_{(2)}^{2}\, , \\
& & \\
F_{\chi\phi} & = & -2 R_{2} {\rm ch}\chi\, ,\\
& & \\
F_{\theta\varphi} & = & 2R_{2}\sin{\theta}\, ,\\
  \end{array}
\right.
\end{equation}

\noindent with 

\begin{equation}
\left\{
  \begin{array}{rcl}
d\Pi_{(2)}^{2} & \equiv & {\rm ch}^{2}\chi\, d\phi^{2}-d\chi^{2}\, , \\
& & \\
d\Omega_{(2)}^{2} & \equiv & d\theta^{2}+\sin^{2}{\theta}d\varphi^{2}\, ,\\
  \end{array}
\right.
\end{equation}

\noindent The metric is that of the direct product of that of $aDS_{2}$
with radius $R_{2}$ in global coordinates $\phi\in [0,2\pi), \,
\chi\in[0,\infty)$ and that of $S^{2}$ with radius $R_{2}$ in standard
spherical coordinates $\theta\in [0,\pi], \, \varphi\in [0,2\pi)$.  It
is known to be maximally supersymmetric in $N=2,d=4$ supergravity
\cite{Gibbons:1984kp,Kallosh:1992gu} in both the electric and magnetic
cases, since the whole $N=2,d=4$ supergravity is invariant under
chiral/dual transformations.
\subsubsection{Electric Case}
Following the rules found in the previous section (with
$s(\alpha)=+1$) and, further, {\it assuming that the compact
  coordinate} $y\in [0,4\pi R_{2})$ and using instead $\psi=y/R_{2}$,
we find the d=5 solution

\begin{equation}
\label{eq:ads2xs3}
\left\{
  \begin{array}{rcl}
d\hat{s}^{2} & = &  R_{2}^{2}\, d\Pi_{(2)}^{2}
-(2R_{2})^{2}\, d\Omega_{(3)}^{2}\, , \\
& & \\
\hat{\mathcal{F}}_{\chi\phi} & = & \sqrt{3}R_{2}\, {\rm ch}\chi\, ,\\
  \end{array}
\right.  
\end{equation}

\noindent with 

\begin{equation}
d\Omega_{(3)}^{2} \equiv  
{\textstyle\frac{1}{4}}\left[
d\Omega_{(2)}^{2}
+\left(d\psi +\cos{\theta}d\varphi\right)^{2}\right]\, ,
\end{equation}

\noindent which is the direct product of $aDS_{2}$
with radius $R_{2}$ in global coordinates $\phi\in [0,2\pi), \,
\chi\in[0,\infty)$ and that of $S^{3}$ with radius $2 R_{2}$ in
Euler-angle coordinates $\theta\in [0,\pi], \, \varphi\in [0,2\pi), \,
\psi\in[0,4\pi)$. This solution is the near-horizon limit of a
5-dimensional extreme black hole and it is maximally supersymmetric
\cite{Chamseddine:1996pi}.


\subsubsection{Magnetic Case}

Straightforward application of the oxidation rules, now with the
compact coordinate in the range $y\in [0,2\pi R_{2})$ with $\eta
=y/R_{2}$ and rescaling $\chi \to \chi/2$ leads us to

\begin{equation}
\label{eq:ads3xs2}
\left\{
  \begin{array}{rcl}
d\hat{s}^{2} & = &  (2R_{2})^{2}\, d\Pi_{(3)}^{2} 
-R_{2}^{2}\,d\Omega_{(2)}^{2}\, , \\
& & \\
\hat{\mathcal{F}}_{\theta\varphi} & = & -\sqrt{3}R_{2}\, \sin{\theta}\, ,\\
  \end{array}
\right.  
\end{equation}

\noindent with 

\begin{equation}
 d\Pi_{(3)}^{2} \equiv {\textstyle\frac{1}{4}}\left[d\Pi_{(2)}^{2}
-\left(d\psi +{\rm sh}(\chi/2)d\phi\right)^{2}  \right]\, ,
\end{equation}

\noindent the metric of $aDS_{3}$ in a form that suggests that this
spacetime can be understood as an $S^{1}$ fibration over $aDS_{2}$.

The above 5-dimensional solution has the metric of $aDS_{3}\times
S^{2}$ which is the near-horizon limit of the extreme $d=5$ string
\cite{Gibbons:1994vm}.

It has been observed many times that $N=2,d=5$ supergravity (its
action and field content) is a theory that resembles very much
$N=1,d=11$ supergravity \cite{Cremmer:1980gs}. One additional
similarity is the presence of two maximally supersymmetric vacua
($aDS_{4}\times S^{7}$ and $aDS_{7}\times S^{4}$) which are
respectively the near-horizon limits of the solutions that describe
the extended objects of the theory: black hole and string in $d=5$ and
M2 and M5 branes in $d=11$. However, it should be clear that we can
obtain new $d=5$ vacua from new $d=4$ vacua, if they exist. As a
matter of fact they do exist: the dyonic RB solutions which have both
electric and magnetic components of the electromagnetic field and
share the same $aDS_{2}\times S^{2}$ metric.


\subsubsection{Dyonic Case}

The dyonic RB solution is given by

\begin{equation}
\left\{
  \begin{array}{rcl}
ds^{2} & = & R_{2}^{2}\,  d\Pi_{(2)}^{2} 
-R_{2}^{2}\, d\Omega_{(2)}^{2}\, , \\
& & \\
F & = &  -{\displaystyle\frac{2}{R_{2}}}\cos{\xi} dr\land dt 
+2R_{2}\sin{\xi}\sin{\theta}d\theta\land d\varphi\, ,\\
  \end{array}
\right.
\end{equation}

\noindent where now, for convenience, we use the following $aDS_{2}$ metric:

\begin{equation}
R_{2}^{2}\,  d\Pi_{(2)}^{2} =
\left(\frac{r}{R_{2}}\right)^{2}dt^{2}
-\left(\frac{R_{2}}{r}\right)^{2}dr^{2}\, .
\end{equation}

This family of solutions, that includes the purely electric and
magnetic cases that we have just seen, has another parameter apart
from the radius $R_{2}$: the duality rotation angle $\xi$. 

Following the oxidation rules we find a 5-dimensional family of
maximally supersymmetric solutions 

\begin{equation}
\label{eq:dyonicuplifting1}
\left\{
  \begin{array}{rcl}
d\hat{s}^{2} & = & 
{\displaystyle\left(\frac{r}{R_{2}}\right)^{2}dt^{2}
-\left(\frac{R_{2}}{r}\right)^{2}dr^{2}
-\left(dy -\frac{r}{R_{2}}\sin{\xi} dt 
+R_{2}\cos{\xi}\cos{\theta}d\varphi\right)^{2}
-R_{2}^{2}d\Omega_{(2)}^{2}}\, ,\\
& & \\
\hat{\mathcal{F}} & = &  
{\displaystyle\frac{\sqrt{3}}{R_{2}}}\cos{\xi}\ dr\land dt 
-\sqrt{3}R_{2}\sin{\xi}\sin{\theta}\ d\theta\land d\varphi\, ,\\
  \end{array}
\right.
\end{equation}

\noindent The explicit form of the Killing spinors in this case reads

\begin{equation}
  \epsilon \;=\; \exp\left(
                    -X\ \log\left( r\right)
                 \right)
                 \exp\left(
                    t\ Y
                 \right)
                 \exp\left(
                    \theta\ Z
                 \right)
                 \exp\left(
                    -\textstyle{\frac{\varphi}{2}}\ \gamma^{34}
                 \right)
                 \epsilon_{(0)} \; ,
\end{equation}
where
\begin{eqnarray}
  X &=& \textstyle{\frac{1}{2}}\left[
           \sin\left(\xi\right)\ \gamma^{02}\ +\ 
           \cos\left(\xi\right)\ \gamma^{0}
        \right] \; , \\
 & & \nonumber \\
  Y &=& \textstyle{\frac{1}{2}}\left[
           \sin\left(\xi\right)\ \gamma^{12}\ +\ 
           \cos\left(\xi\right)\ \gamma^{1}\ -\
           \gamma^{01}
        \right] \; , \\
 & & \nonumber \\
  Z &=& \textstyle{\frac{1}{2}}\left[
           \cos\left(\xi\right)\ \gamma^{24}\ +\ 
           \sin\left(\xi\right)\ \gamma^{4}
        \right] \; ,
\end{eqnarray}

After the coordinate redefinitions

\begin{equation}
\cos\left(\xi\right)\, t\to t\, ,
\hspace{.5cm}
\frac{y}{R_{2}\cos\left(\xi\right)}\to \psi\, ,  
\end{equation}

\noindent takes the form

\begin{equation}
\label{eq:GMT_solution}
\left\{
  \begin{array}{rcl}
d\hat{s}^{2} & = & {\displaystyle\left[\frac{r}{R_{2}}dt 
+R_{2}\sin{\xi}\,\left(d\psi +\cos{\theta}d\varphi \right) \right]^{2} 
-\left(\frac{R_{2}}{r}\right)^{2}dr^{2}
-(2R_{2})^{2}d\Omega_{(3)}^{2}}\, ,\\
& & \\
\hat{\mathcal{F}} & = &  
{\displaystyle\frac{\sqrt{3}}{R_{2}}}\ dr\land dt 
-\sqrt{3}R_{2}\sin{\xi}\sin{\theta}\ d\theta\land d\varphi\, .\\
  \end{array}
\right.
\end{equation}

If we set $R_{2}=1/2$, $\sin{\xi}=j$, $2t\to t$ and $r\to r^2$
we recover a
solution that describes the near-horizon limit of the supersymmetric
\cite{Kallosh:1996vy} rotating $d=5$ black hole, given in
\cite{Gauntlett:1998fz,Townsend:1998ci}. While it was known that in
the zero-rotation limit $j=0$ this solution has the metric of
$aDS_{2}\times S^{3}$, the result in the limiting case $j\to 1$ was 
unknown since it is a singular limit. However, by means of the 
inverse of the above coordinate transformations, the limit can be
taken in such a  way that the limiting metric, at $\xi =\pi /2$, is
regular: $aDS_{3}\times S^{2}$.
Thus, the near-horizon
limit of the $j=1$ supersymmetric rotating black hole and the
near-horizon limit of the string are identical.
\par
{}Finally, let us mention the superalgebras associated to these  
vacua. As was pointed out in Ref. \cite{Gauntlett:1998fz}, the
superalgebra associated to the solution (\ref{eq:GMT_solution}),
is $su(1,1|2)\oplus u(1)$ when $0<j<1$ and gets enhanced to 
$su(1,1|2)\oplus su(2)$ when $j=0$. Combining this with the smooth
$\xi =\pi /2$ limit for the family (\ref{eq:dyonicuplifting1}), one 
sees that the superalgebra associated to the $aDS_{3}\times S^{2}$ has to
be $su(1,1|2)\oplus sl(2,\mathbb{R})$ and not $sl(2,\mathbb{R})\rtimes su(1,1|2)$
as was hinted at in Ref. \cite{Fujii:1998tc}.  
\subsection{Oxidation to $d=6$}
The oxidation of Eq.~(\ref{eq:dyonicuplifting1}) gives, after rotation
of the two isometric coordinates $y,w$ by the angle associated to the
4-dimensional electric-magnetic duality $\xi$

\begin{equation}
\left\{
  \begin{array}{rcl}
w & = & \cos{\xi}\eta +R_{2}\sin{\xi} \psi\, ,\\
& & \\
y & = & -\sin{\xi}\eta +R_{2}\cos{\xi}\psi\, ,\\   
  \end{array}
\right.
\end{equation}

\noindent one recovers the solution

\begin{equation}
\label{eq:ads3xs3}
\left\{
  \begin{array}{rcl}
d\hat{s}^{2} & = &  (2R_{2})^{2}\, d\Pi_{(3)}^{2} 
-(2R_{2})^{2}\,d\Omega_{(3)}^{2}\, , \\
& & \\
\hat{B}^{-} & = & {\displaystyle\frac{r}{R_{2}}}d\eta\land dt
-R_{2}^{2}\cos{\theta}d\varphi\land d\psi\, ,\\
  \end{array}
\right.  
\end{equation}

\noindent whose metric is that of $aDS_{3}\times S^{3}$, the maximally
supersymmetric solution which is the near-horizon limit of a
self-dual string \cite{Gibbons:1994vm}. It is known that the uplifting
of the near-horizon limit of the rotating $d=5$ black hole gives, for
any value of the rotation parameter, $aDS_{3}\times S^{3}$
\cite{Cvetic:1998xh}.
\section{Conclusions}
\label{sec-conclusions}
In this article we have shown that the known supersymmetric vacua 
of the $d=6$ $N=(2,0)$, $d=5$ $N=2$ and $d=4$ $N=2$ supergravity are
linked by dimensional reduction. Although this may come as a bit of a 
surprise when thinking in terms of dimensional reduction, it is quite 
obvious from the oxidation point of view: since all three theories have
8 supercharges and oxidation cannot reduce the number of preserved 
supersymmetries, a lower dimensional maximally supersymmetric solution
must lift to a maximally supersymmetric solution.
\par
{}From the supergravity point of view, the relations can hold because
the dimensionally reduced theories can be truncated consistently
to the minimal $N=2$ supergravity, {\em i.e.} without any matter
couplings.
A subtle point in the dimensional reduction is that for the 
Killing spinors to survive the dimensional reduction, the Killing
spinors must be independent of the compact coordinates. 
In a coordinate independent way, this means that there must be a Killing
vector whose action on the Killing spinor vanishes, or put differently,
there is a bosonic generator in the superalgebra associated to the 
solution \cite{Gauntlett:1998kc}, that is represented trivially on
the supercharges.
Actually, it is not difficult to see that from the superalgebra point
of view, the relation between the $N=2$ vacua was going to hold.
\par
{}For definiteness let us consider the superalgebras associated to the 
$aDS_{p}\times S^{q}$ spacetimes (See table (\ref{tabel_1})),
the analogous results for the KG-waves being obtainable by a  
In{\"o}n{\"u}-Wigner contraction on their $aDS\times S$ counterparts
\cite{Hatsuda:2002kx}.\footnote{The exception is of course the family
of metrics in Eq. (\ref{eq:dyonicuplifting1}),
when $\xi\neq 0,\pi /2$, since its Penrose contraction has 2 more isometries.}
\begin{table}
\begin{center}
\begin{tabular}{|c|c|c|c|}
\hline
Space & Theory & Solution & Superalgebra \\
\hline
$aDS_{3}\times S^{3}$ & $N=(2,0)$ $d=6$ &(\ref{eq:ads3xs3}) & $ su(1,1|2)\oplus sl(2,\mathbb{R})\oplus su(2) $ \\
\hline
$aDS_{3}\times S^{2}$ & $N=2$ $d=5$ & (\ref{eq:ads3xs2}) & $ su(1,1|2)\oplus sl(2,\mathbb{R})$ \\
\hline
$aDS_{2}\times S^{3}$ & $N=2$ $d=5$ & (\ref{eq:ads2xs3}) & $ su(1,1|2)\oplus su(2) $ \\[.1cm]
\hline
Dyonic  & $N=2$ $d=5$ & (\ref{eq:dyonicuplifting1}) & $ su(1,1|2)\oplus u(1) $ \\[.1cm]
\hline
$aDS_{2}\times S^{2}$ & $N=2$ $d=4$ &(\ref{eq:ads2xs2}) & $ su(1,1|2) $ \\[.1cm]
\hline
\end{tabular}
\caption{\label{tabel_1} Solutions and their associated superalgebras.}
\end{center}
\end{table}
It is clear that the way to preserve supersymmetry is by embedding
the generator of translations in the compactification direction, 
in the non-$su(1,1|2)$
part of the superalgebra. For the dimensional reduction from $d=6$ to $d=5$,
there are basically
3 choices, corresponding to the 3 5-dimensional solutions given in
Eqs. (\ref{eq:ads2xs3},\ref{eq:ads3xs2},\ref{eq:dyonicuplifting1}).
For a further reduction to $d=4$ there is basically one way to embed such a
translation generator.
Note that the chain of relations exposed in this letter is quite
unique among the vacua:
Had we considered for example the $aDS_{3}\times S^{3}$ solution in the 
$d=6$ $N=(4,0)$ supergravity, we would have had to conclude that, since the 
associated superalgebra is $su(1,1|2)\oplus su(1,1|2)$, there is no way
to preserve the 16 supercharges in a circle compactification. \\ 

{\bf Note added in proof:} After this paper was accepted for publication, 
the preprint \cite{Gauntlett:2002nw} appeared, giving a complete 
clasification of all supersymmetric solutions of $d=5$, $N=2$ supergravity. 
In that reference, a new maximally supersymmetric solution (a five 
dimensional generalization of the G\"odel universe) is found.

\section*{Acknowledgments}
The authors would like to thank N.~Alonso-Alberca, E.~Bergshoeff, P.~Bain
and M.~Zamaklar for useful discussions. T.O. would like to thank the 
Newton Institute for its hospitality and financial support and 
M.M.~Fern{\'a}ndez for her continuous support. This work was supported in part
by the E.U. RTN programme HPRN-CT-2000-00148. 
The work of E.L.-T. and T.O. is supported by the Spanish grant FPA2000-1584.
\appendix

%
%

\begin{thebibliography}{30}
%
%
\bibitem{Kowalski-Glikman:wv}
J.~Kowalski-Glikman,
Phys.\ Lett.\ B {\bf 134} (1984) 194.
%
\bibitem{Figueroa-O'Farrill:2001nz}
J.~Figueroa-O'Farrill and G.~Papadopoulos,
JHEP {\bf 0108} (2001) 036
[arXiv:hep-th/0105308].
%
\bibitem{Blau:2001ne}
M.~Blau, J.~Figueroa-O'Farrill, C.~Hull and G.~Papadopoulos,
JHEP {\bf 0201} (2002) 047
[arXiv:hep-th/0110242].
%
\bibitem{Gueven:ad}
R.~Gueven,
Phys.\ Lett.\ B {\bf 191} (1987) 275;
D.~Amati and C.~Klimcik,
Phys.\ Lett.\ B {\bf 219} (1989) 443;
H.~J.~de Vega and N.~Sanchez,
Phys.\ Lett.\ B {\bf 244} (1990) 215;
G.~T.~Horowitz and A.~R.~Steif,
Phys.\ Rev.\ Lett.\  {\bf 64} (1990) 260;
Phys.\ Rev.\ D {\bf 42} (1990) 1950;
E.~A.~Bergshoeff, R.~Kallosh and T.~Ort\'{\i}n,
Phys.\ Rev.\ D {\bf 47} (1993) 5444
[arXiv:hep-th/9212030];
C.~R.~Nappi and E.~Witten,
Phys.\ Rev.\ Lett.\  {\bf 71} (1993) 3751
[arXiv:hep-th/9310112];
O.~Jofre and C.~Nunez,
Phys.\ Rev.\ D {\bf 50} (1994) 5232
[arXiv:hep-th/9311187];
A.~A.~Kehagias and P.~Meessen,
Phys.\ Lett.\ B {\bf 331} (1994) 77
[arXiv:hep-th/9403041];
K.~Sfetsos and A.~A.~Tseytlin,
Nucl.\ Phys.\ B {\bf 427} (1994) 245
[arXiv:hep-th/9404063];
E.~Kiritsis, C.~Kounnas and D.~Lust,
Phys.\ Lett.\ B {\bf 331} (1994) 321
[arXiv:hep-th/9404114];
A.~A.~Tseytlin,
Class.\ Quant.\ Grav.\  {\bf 12} (1995) 2365
[arXiv:hep-th/9505052];
%
\bibitem{art:RR_quant}
R.R.~Metsaev,
Nucl.\ Phys.\ B {\bf 625} (2002) 70
[arXiv:hep-th/0112044];
R.R.~Metsaev and A.A.~Tseytlin,
Phys.\ Rev.\ D {\bf 65} (2002) 126004
[arXiv:hep-th/0202109];
J.G.~Russo and A.A.~Tseytlin,
JHEP {\bf 0204} (2002) 021
[arXiv:hep-th/0202179].
%
\bibitem{art:D_branes}
M.~Billo and I.~Pesando,
Phys.\ Lett.\ B {\bf 536} (2002) 121
[arXiv:hep-th/0203028];
A.~Dabholkar and S.~Parvizi,
arXiv:hep-th/0203231.
A.~Kumar, R.~R.~Nayak and Sanjay,
arXiv:hep-th/0204025.
P.~Bain, P.~Meessen and M.~Zamaklar,
arXiv:hep-th/0205106.
%
\bibitem{Skenderis:2002vf}
K.~Skenderis and M.~Taylor,
arXiv:hep-th/0204054.
%
%
\bibitem{kn:Pen6} R.~Penrose,
                  in
                  {\sl Differential Geometry and Relativity}, Reidel, 
                  Dordrecht (1976) 271-275.
%
\bibitem{Gueven:2000ru}
R.~Gueven,
Phys.\ Lett.\ B {\bf 482} (2000) 255
[arXiv:hep-th/0005061].
%
\bibitem{Blau:2002dy}
M.~Blau, J.~Figueroa-O'Farrill, C.~Hull and G.~Papadopoulos,
arXiv:hep-th/0201081.
%
\bibitem{Blau:2002rg}
M.~Blau, J.~Figueroa-O'Farrill and G.~Papadopoulos,
arXiv:hep-th/0202111.
%
\bibitem{art:hpp_cft}
D.~Berenstein, J.M.~Maldacena and H.~Nastase,
JHEP {\bf 0204} (2002) 013
[arXiv:hep-th/0202021];
N.~Itzhaki, I.R.~Klebanov and S.~Mukhi,
JHEP {\bf 0203} (2002) 048
[arXiv:hep-th/0202153];
J.~Gomis and H.~Ooguri,
arXiv:hep-th/0202157;
U.~Gursoy, C.~Nunez and M.~Schvellinger,
JHEP {\bf 0206} (2002) 015
[arXiv:hep-th/0203124].
S.R.~Das, C.~Gomez and S.J.~Rey,
arXiv:hep-th/0203164;
D.~Brecher, C.V.~Johnson, K.J.~Lovis and R.C.~Myers,
arXiv:hep-th/0206045.
%
\bibitem{Duff:1998us}
M.J.~Duff, H.~Lu and C.~N.~Pope,
Nucl.\ Phys.\ B {\bf 532} (1998) 181
[arXiv:hep-th/9803061].
%
\bibitem{Michelson:2002wa}
J.~Michelson,
arXiv:hep-th/0203140.
%
\bibitem{Gheerardyn:2001jj}
J.~Gheerardyn and P.~Meessen,
Phys.\ Lett.\ B {\bf 525} (2002) 322
[arXiv:hep-th/0111130].
%
\bibitem{Bergshoeff:1994cb}
E.~Bergshoeff, R.~Kallosh and T.~Ort\'{\i}n,
Phys.\ Rev.\ D {\bf 51} (1995) 3009
[arXiv:hep-th/9410230].
%
\bibitem{Bakas:1994ba}
I.~Bakas,
Phys.\ Lett.\ B {\bf 343} (1995) 103
[arXiv:hep-th/9410104].
%
\bibitem{Duff:1997qz}
M.~J.~Duff, H.~Lu and C.~N.~Pope,
Phys.\ Lett.\ B {\bf 409} (1997) 136
[arXiv:hep-th/9704186].
%
\bibitem{Tod:pm}
K.~p.~Tod,
Phys.\ Lett.\ B {\bf 121} (1983) 241.
%
\bibitem{Gauntlett:2002nw}
J.~P.~Gauntlett, J.~B.~Gutowski, C.~M.~Hull, S.~Pakis and H.~S.~Reall,
arXiv:hep-th/0209114.
%
\bibitem{Meessen:2001vx}
P.~Meessen,
Phys.\ Rev.\ D {\bf 65} (2002) 087501
[arXiv:hep-th/0111031].
%
\bibitem{Gibbons:1994vm}
G.~W.~Gibbons, G.~T.~Horowitz and P.~K.~Townsend,
Class.\ Quant.\ Grav.\  {\bf 12} (1995) 297
[arXiv:hep-th/9410073].
%
\bibitem{Kallosh:1996vy}
R.~Kallosh, A.~Rajaraman and W.~K.~Wong,
Phys.\ Rev.\ D {\bf 55} (1997) 3246
[arXiv:hep-th/9611094].
%
\bibitem{Gauntlett:1998fz}
J.~P.~Gauntlett, R.~C.~Myers and P.~K.~Townsend,   
Class.\ Quant.\ Grav.\  {\bf 16} (1999) 1
[arXiv:hep-th/9810204].
%
\bibitem{Kowalski-Glikman:1985im}
J.~Kowalski-Glikman,
Phys.\ Lett.\ B {\bf 150} (1985) 125.
%
\bibitem{kn:Rob} I.~Robinson,
                 {\it Bull.~Acad.~Polon.~Sci.}~\textbf{7}(1959), 351;
B.~Bertotti,
                  {\it Phys.~Rev.}~\textbf{116} (1959) 1331.
%
\bibitem{Boonstra:1998yu}
H.~J.~Boonstra, B.~Peeters and K.~Skenderis,
Nucl.\ Phys.\ B {\bf 533} (1998) 127
[arXiv:hep-th/9803231].
\bibitem{Gibbons:1999uv}
G.~W.~Gibbons and C.~A.~Herdeiro,
Class.\ Quant.\ Grav.\  {\bf 16} (1999) 3619
[arXiv:hep-th/9906098].
%
\bibitem{kn:CaWa} M.~Cahen and N.~Wallach,
                  {\it Bull.~Am.~Math.~Soc.}~\textbf{76} (1970) 585-591.
%
\bibitem{Papadopoulos:1999tg}
G.~Papadopoulos,
JHEP {\bf 9904} (1999) 014
[arXiv:hep-th/9902166];
M.~Cvetic, H.~Lu and C.~N.~Pope,
arXiv:hep-th/0203229;
arXiv:hep-th/0203082;
J.~P.~Gauntlett and C.~M.~Hull,
arXiv:hep-th/0203255;
I.~Bena and R.~Roiban,
arXiv:hep-th/0206195.
%
\bibitem{Nishino:dc}
H.~Nishino and E.~Sezgin,
Nucl.\ Phys.\ B {\bf 278} (1986) 353.
%
\bibitem{Meessen:1998qm}
P.~Meessen and T.~Ort\'{\i}n,
Nucl.\ Phys.\ B {\bf 541} (1999) 195
[arXiv:hep-th/9806120].
%
\bibitem{Cremmer:1980gs}
E.~Cremmer,
in {\it C80-06-22.1.1}
LPTENS 80/17
{\it Invited paper at the Nuffield Gravity Workshop, Cambridge, Eng., Jun 22 - Jul 12, 1980}.
%
\bibitem{Chamseddine:1980sp} A.~H.~Chamseddine and H.~Nicolai,
Phys.\ Lett.\ B {\bf 96} (1980) 89.
%
\bibitem{kn:MaPa} S.D.~Majumdar,
                {\it Phys.~Rev.}~\textbf{72} (1947) 390-398;
A.~Papapetrou,
               {\it Proc.~Roy.~Irish.~Acad.}~\textbf{A51} (1947) 191.

%
\bibitem{Gibbons:1984kp}
G.~W.~Gibbons,
Print-85-0061 (CAMBRIDGE)
{\it Three lectures given at GIFT Seminar on Theoretical Physics, San Feliu de Guixols, Spain, Jun 4-11, 1984}.
%
\bibitem{Kallosh:1992gu}
R.~Kallosh,
Phys.\ Lett.\ B {\bf 282} (1992) 80
[arXiv:hep-th/9201029].
%
\bibitem{Chamseddine:1996pi}
A.~H.~Chamseddine, S.~Ferrara, G.~W.~Gibbons and R.~Kallosh,
Phys.\ Rev.\ D {\bf 55} (1997) 3647
[arXiv:hep-th/9610155].
%
\bibitem{Townsend:1998ci}
P.~K.~Townsend,
arXiv:hep-th/9901102.
%
\bibitem{Fujii:1998tc}
A.~Fujii, R.~Kemmoku and S.~Mizoguchi,
Nucl.\ Phys.\ B {\bf 574} (2000) 691
[arXiv:hep-th/9811147].
%
\bibitem{Cvetic:1998xh}
M.~Cvetic and F.~Larsen,
Nucl.\ Phys.\ B {\bf 531} (1998) 239
[arXiv:hep-th/9805097].
%
\bibitem{Gauntlett:1998kc}
J.P.~Gauntlett, R.C.~Myers and P.K.~Townsend,
Phys.\ Rev.\ D {\bf 59} (1999) 025001
[arXiv:hep-th/9809065].
%
\bibitem{Hatsuda:2002kx}
M.~Hatsuda, K.~Kamimura and M.~Sakaguchi,
arXiv:hep-th/0204002.

%
%
\end{thebibliography}
\end{document}